\documentclass[twocolumn,secnumarabic,amssymb, nobibnotes, aps, prb,superscriptaddress]{revtex4-1}

\newcommand{\macro}[1]{\texttt{\textbackslash#1}}
\newcommand{\m}[1]{\macro{#1}}

\usepackage{graphicx}
\usepackage{dcolumn}
\usepackage{bm}
\usepackage{hyperref}
\usepackage[]{natbib}
\usepackage{amsmath}
\usepackage{amsfonts}

\begin{document}
	
	\preprint{APS/123-QED????????????}
	
	\title{Change of Corner Charge and Adiabatic Current Distribution in Two Dimensional Insulators with Inversion Symmetry}
	
	\author{Xuzhe Ying}
	\affiliation{Department of Physics and Astronomy, University of Waterloo, Waterloo, ON, N2L 3G1, Canada}


	
	
	\begin{abstract}
		We discuss the change of the corner charge for noninteracting two dimensional insulators with inversion symmetry undergoing adiabatic evolution. We show that the change of the corner charge is accounted for by the adiabatic current flowing along the edges of the system. The study of systems with quasi-1D geometry is necessary to derive the analytical expression for the adiabatic current. This fact suggests that the change of the corner charge is neither a purely bulk nor edge effect, but rather a mixed one. The derived adiabatic current was examined and shows good agreement with the numerical calculation of Benalcazar-Bernevig-Hughes model. 
		\begin{description}
			\item[PACS numbers]
		\end{description}
	\end{abstract}
	
	\pacs{1111}
	\maketitle
	
	\section{Introduction}
	\label{sec:intro}	
	
	The proposal of higher order topological insulator (HOTI)\cite{Benalcazar61,PhysRevB.96.245115,PhysRevB.97.094508,PhysRevB.97.205136,PhysRevB.98.235102,PhysRevResearch.2.022049,PhysRevResearch.2.033029,ni2020demonstration,schindler2018higher,parameswaran2017topological} has revived the wide interest for the electric multipole moments in crystalline insulators\cite{PhysRevB.103.035147,PhysRevResearch.2.043012,PhysRevB.92.041102,PhysRevB.100.245133,PhysRevB.100.245134,PhysRevB.102.165120,PhysRevB.102.235149,PhysRevB.103.125129,10.21468/SciPostPhys.10.4.092,huang2021effective,dubinkin2020higher,may2021crystalline}. Although electric multipole moments are fundamental concepts in electromagnetism, the search for a self-consistent quantum mechanical theory is surprisingly challenging and brings about a change of paradigm\cite{RevModPhys.66.899,resta2007theory,vanderbilt2018berry,PhysRevB.47.1651,PhysRevB.48.4442,PhysRevLett.80.1800}.
	
	A classical description of polarization in (crystalline) insulators is based on the Clausius-Mossotti (CM) model\cite{vanderbilt2018berry}. In CM model, localized and nonoverlapping charge distribution is assumed. Polarization is defined as dipole moment per unit cell. Quantum mechanics completely altered the paradigm. In crystalline materials, electrons are described by Bloch wavefunctions extending over the whole sample\cite{ashcroft1976solid}. The classical CM description of polarization becomes problematic. 
	
	The modern theory of polarization based on quantum mechanics was developed in the recent decades\cite{RevModPhys.66.899,resta2007theory,vanderbilt2018berry,PhysRevB.47.1651,PhysRevB.48.4442,PhysRevLett.80.1800}. Instead of polarization itself, the change of polarization was proposed to be the physical observable and is measured by the change of surface charge\cite{resta2007theory}. Namely, the change of polarization is captured by a bulk adiabatic current pumping charge to the boundary of the material\cite{PhysRevB.47.1651}. The relation between the bulk polarization and the surface charge is sometimes termed as bulk-boundary correspondence\cite{PhysRevB.48.4442,PhysRevB.95.035421}. On the technical aspect, the modern theory of polarization naturally introduces the concepts of Berry connection and Berry curvature\cite{vanderbilt2018berry,RevModPhys.66.899}, which underlies physics of topological insulators\cite{RevModPhys.82.3045,shen2012topological,bernevig2013topological,PhysRevB.59.14915}.
	
	The study of HOTI and higher electric multipole moment put much stronger challenge. The first proposal for HOTI is known as Benalcazar-Bernevig-Hughes (BBH) model\cite{Benalcazar61,PhysRevB.96.245115}. BBH model is a two dimensional generalization of Su-Schrieffer-Heeger model\cite{PhysRevB.22.2099}. With proper sample geometry (square), BBH model supports degenerate zero energy corner states. More profoundingly, the charge accumulated at the corners were found to be half quantized, which is a result of lattice symmetry of the model, e.g., reflection, inversion and rotation symmetry\cite{Benalcazar61}. The authors argued that the same symmetries also require the quadrupole moment to be quantized.
	
	Motivated by HOTI, many attempts have been devoted to understanding quadrupole moments in solids with generic crystal symmetry\cite{PhysRevB.103.035147,PhysRevB.102.235149,PhysRevResearch.2.043012,PhysRevB.100.245133,PhysRevB.100.245134,PhysRevB.102.165120,PhysRevB.102.235149}. The focus was primarily on the bulk defintion of quadrupole moment and the prediction of corner charge in a generic situation. Naturally, a relation between bulk quadrupole moment, edge polarization and the corner charge was proposed\cite{PhysRevB.96.245115}. However, the task turns out to be quite subtle and challenging\cite{PhysRevB.100.245133}. In pursuit of an equation to predict the corner charge, the authors of Ref.~[\onlinecite{PhysRevB.103.035147}] realized that any definition of bulk quadrupole moment or edge polarization is gauge depenedent. Nevertheless, the corner charge is a gauge independent quantity. In addition, previous works found that a study of quasi-1D systems is necessary to predict the corner charge\cite{PhysRevResearch.2.043012,PhysRevB.103.035147}. This fact was called the bulk-and-edge to corner correspondence\cite{PhysRevResearch.2.043012}. 
	
	In this work, we focus on the change of the corner charge as well as the associated adiabatic edge current. This work provides a different perspective on the prediction of corner charge, supplementing previous works. The current approach is based on the adiabatic evolution of an insulating system, similar to the modern theory of polarization where the adiabtic current is a necessary constituent\cite{PhysRevB.47.1651,RevModPhys.66.899}. The goal of this work is two fold. First, an equation to predict the adiabatic current and the change of corner charge is proposed; Second, we address that there is no obvious separation of the adiabatic current into purely edge or purely bulk contributions. The second point is in accordance with previous works, suggesting that edge polarization or bulk quadrupole moment are not independently defined\cite{PhysRevB.103.035147}. Instead, the corner charge remains a well defined physical quantity.
	
	
	To demonstrate the main ideas, we consider a 2D insulator with inversion symmetry. A finite size sample is separated into various macroscopic regions - bulk, edges and corners, as shown in Fig.~\ref{Fig:SystemDemo}. We took square lattice as an example.  Due to the inversion symmetry, instead of the edges, the corner regions may support excess charge, compared to the bulk of the system. To study the slow change of the corner charge, we considered quasi-1D systems (as shown in Fig.~\ref{Fig:SystemDemoQuasi1D}) in both $x$- and $y$-direction. We derived the edge adiabatic current by solving the time-dependent Schrodinger equation within quasi-1D geometry. We show that the adiabatic current is gauge independent and is indeed localized at the edges. Then, we examined the result of the adiabatic current for BBH model, whose lattice structure and spectrum are shown in Fig. \ref{Fig:BBHModel}-\ref{Fig:BBH1DSpectrum}. Fig.~\ref{Fig:CC_Current} shows a good agreement between the proposed equation of adiabatic current and the numerical calculation of the change of corner charge, justifying the derived equation of the adiabatic current.
	
	The rest of the article is organized as follows: Sec.~\ref{Sec:Model_CornerCharge} introduces the general model and definition of the corner charge; Sec.~\ref{Sec:AdiabaticCurrent} outlines the derivation and discusses key properties of the adiabatic current; Sec.~\ref{Sec:BBH_Model} applies the developed theory of adiabatic current to BBH model; Sec.~\ref{Sec:Conclusion} concludes this article with furthur discussions. This article is supplemented with an appendix on certain technical details: Appendix.~\ref{Sec:Observables} validates the observables constructed in Sec.~\ref{Sec:AdiabaticCurrent}; Appendix.~\ref{Sec:DetailsOnTB} provides the spectrum and wavefunctions of BBH model, which are necessary to calculate the abiabatic current.

	\section{General Model and Corner Charge}
	\label{Sec:Model_CornerCharge}
	
	\begin{figure}[tb]
		\centering
		\includegraphics[scale=0.36]{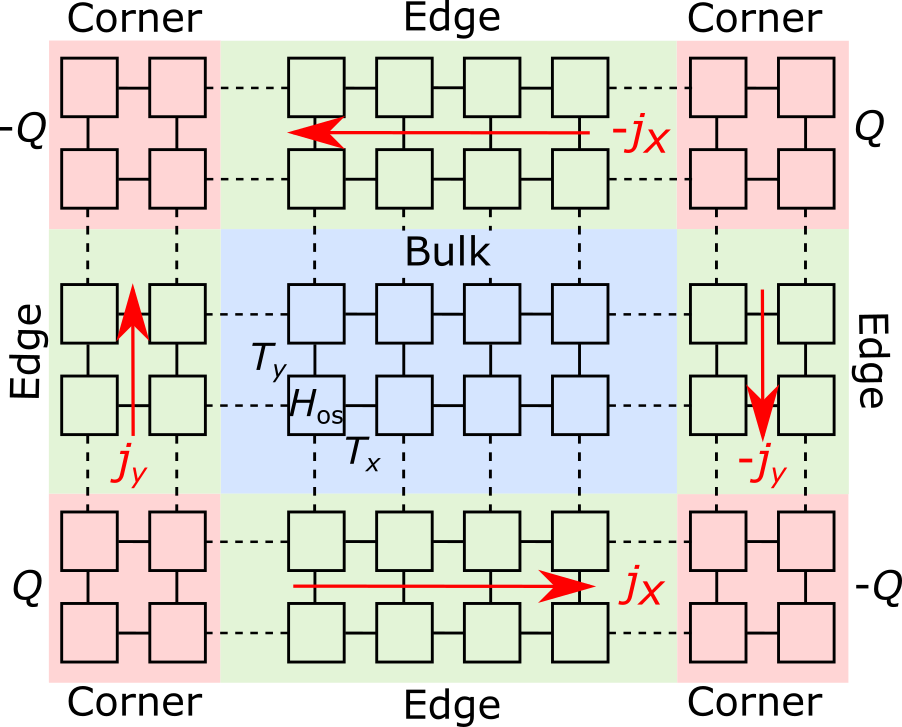}
		\caption{Schematic lattice structure for a finite size system. Each box represents a Wigner-Seitz unit cell; each line represents inter-unit-cell bond. The system is separated into three macroscopic regions: corner (red), edge (green) and bulk (blue). All three regions are large compared to the size of the unit cell.}
		\label{Fig:SystemDemo}
	\end{figure}
	
	In this article, we consider spinless, noninteracting two dimensional insulators with inversion symmetry. Without loss of generality, we assume that the unit cells form a square lattice as an example, Fig.~\ref{Fig:SystemDemo}. Each unit cell contains multiple sublattices. We assume the system to be at integer filling, such that the lower $N$ bands are fully filled while the upper $M$ bands are fully empty.
	
	Due to the presumed inversion symmetry, the bulk polarization may take discrete values, quantized to half integers\cite{resta2007theory}. We focus on the case of $(p_x,p_y)=(0,0)$ so that the edges do not hold excess charge. In other cases, gapless edge states may appear just like SSH model\cite{PhysRevB.22.2099,PhysRevLett.62.2747}, where the corner charge and the adiabatic edge current becomes ill-defined. Thus, in this article, we will assume the quasi-1D system to be insulating.
	
	For a finite size system, we may separate it into several macroscopic regions as schematically shown in Fig.~\ref{Fig:SystemDemo}. Those regions are the corners (red), edges (green) and the bulk (blue). For spinless electrons, the bulk electron number per unit cell is $N$, given that in the bulk the lower $N$ bands are assumed to be filled. The edges should have the same electron density as the bulk, following the discussion of inversion symmetry and polarization in the previous paragraph.
	
	The corners of the sample may hold excess charges at equilibrium. As shown in Fig.~\ref{Fig:SystemDemo}, the bottom-left and the top-right corner holds the same charge $Q$, while the top-left and the bottom-right corner holds $-Q$. There are multiple ways to define the corner charge\cite{vanderbilt2018berry,PhysRevB.103.035147}. We expect the difference between different definitions to be well controlled for large system and corner size in approaching the thermodynamic limit. Here, for simplicity, we took a heuristic defintion as follows:
	\begin{equation}
		Q_{\text{c}}=\sum_{\text{Occupied States}}\sum_{\boldsymbol{R}\in\text{Corner}}\psi^{\dagger}_i(\boldsymbol{R})\psi_i(\boldsymbol{R})-Q_{\text{ref}}
	\end{equation}
	where $\psi_i(\boldsymbol{R})$ is the wavefunction for energy level $i$; the summation is over all the occupied states and one of the four corner regions. The reference charge $Q_{\text{ref}}$ may be chosen such that the corner charge vanishes in certain situation, e.g., the trivial phase of BBH model to be introduced in Sec.~\ref{Sec:BBH_Model}. Equivalently, one may use the bulk charge density as a reference. Practically, the definition of corner charge above converges for corners  with a side length of a few unit cell layers ($\sim 10$). 
	
	Instead of direct prediction, we ask the question of how the corner charge changes if the systems is slowly deformed. A knowledge of the change of the corner charge can provide an accurate prediction of the corner charge.
	
	If the system is subject to slow adiabatic  evolution, the corner charge will vary in time. As shown in Fig.~\ref{Fig:SystemDemo}, the change of the corner charge is associated with the adiabatic current along the edges with the following relation:
	\begin{equation}
		-\frac{dQ}{dt}=j_x+j_y
	\end{equation}
	Moreover, the adiabatic current flows in the opposite directions for bottom and top edges as well as the left and the right edges. The bulk of the system is free from any macroscopic current. This is in accordance with the assumption that the system remains dipole free in all time.
	
	Generically, the edge current may contain two parts. The first part corresponds to the itinerant boundary current, originated from the orbital magnetic moment\cite{vanderbilt2018berry,RevModPhys.82.1959,PhysRevB.100.054408,PhysRevLett.95.137204,PhysRevLett.95.137205}. This is an equilibrium current. Thus, it does not lead to any charge accumulation. More interestingly is the second part associated with the slow adiabatic evolution of the system. 
	
	In the rest of the article, we show that the adiabatic part of the edge current can be calculated in a quasi-1D setup, as shown in Fig.~\ref{Fig:SystemDemoQuasi1D}. A study of quasi-1D geometry with both orientations (horizontal and vertical) is necessary to predict the change of the corner charge accurately. 

	\section{Adiabatic Edge Current from Quasi-1D Geometry}
	\label{Sec:AdiabaticCurrent}
	
\begin{figure}[tb]
		\centering
		\includegraphics[scale=0.35]{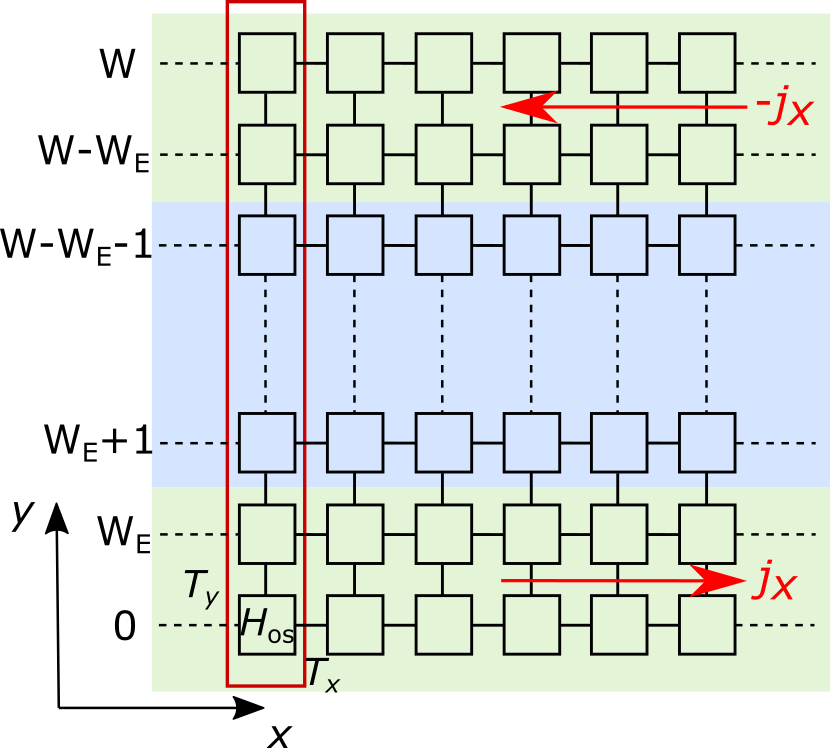}
		\caption{Schematic lattice structure for a quasi 1D system of width $W$. Red box indicates the quasi 1D unit cell. The (green) edge rigions has width $W_{\text{E}}+1$. The bulk region is indicated by the blue region. In unit of lattice constant, $1\ll W_{\text{E}}\ll W$. }
		\label{Fig:SystemDemoQuasi1D}
	\end{figure}

	In this section, we study the adiabatic evolution of samples with quasi-1D geometry of width $W+1$, shown in Fig.~\ref{Fig:SystemDemoQuasi1D}. We derive the equation for the adiabatic currents and show that the adiabatic current is gauge-independent and localized at the boundary of the sample. The analysis for quasi-1D system in both $x$- and $y$-directions is necessary. In this section, we present the analysis for $x$-direction in detail. The analysis for $y$-direction fully parallels that of $x$-direction.

	\subsection{General Consideration: Quasi-1D Subbands and Wavefunctions}
	
	In this subsection, we introduce the basic setup for the quasi-1D systems and the necessary notations.
	
	As discussed in Sec.~\ref{Sec:Model_CornerCharge}, we focus on the situation with gapped bulk and edge. In this situation, the adiabatic evolution to be discussed in this section are well-defined.
	
	To be more specific, we assume the lower $\tilde{N}$ subbands with energy $E_{n,\kappa_y,k_x}$ are fully filled. Meanwhile, the upper $\tilde{M}$ subbands with $E_{m,\kappa_y,k_x}$ are fully empty. Here, the subbands depends on the following quantum numbers: $k_x$ is the quasi-momentum in the infinite $x$-direction; $\kappa_y$ is the transverse quantum number; lastly, $n/m$ denotes the collection of all other quantum numbers. The transverse quantum number $\kappa_y$ is determined from the quantization condition (or the boundary conditions at the bottom and top edges). Generically, it takes $W+1$ discrete values. The other quantum numbers are assigned the following values: $n=1,2,\cdots N$ and $m=N+1, N+2,\cdots N+M$. This is in accordance with the assumption that the lower $N$ (upper $M$) bulk bands are filled (empty). 
	
	The quasi-1D Bloch wavefunction with quantum number $(n/m, \kappa_y, k_x)$ is assumed to be:
	\begin{equation}
		\psi_{n/m, \kappa_y,k_x}(x,y)=e^{ik_xx}U_{n/m,\kappa_y,k_x}(y)
	\end{equation}
	where $U_{n/m,\kappa_y,k_x}(y)$ is the periodic part of the quasi-1D Bloch wavefunction. It depends on the position $y$ in the transverse direction. Later, we will look into the algebraic structure of $U_{n/m,\kappa_y,k_x}(y)$ to show that the adiabatic current is localized at the boundary.

	\subsection{Adiabatic Evolution and Expectation Value of Observables}
	\label{Sec:AdiaEvo_ExpValObser}
	
	In this subsection, we discuss the adiabatic evolution of the quasi-1D samples and construct the expectation values of observables based on single particle wavefunctions for noninteracting systems. 
	
	Due to the finite energy gap between the filled and empty states, the many body spectrum is gapped. The adiabatic approximation can be applied to study the time evolution of the system with a slowly changing Hamiltonian. Within the adiabatic approximation, the initial condition for the many body Schrodinger equation is inconsequential. Namely, at each instant of time, the wavefunction is primarily dominated by the groundstate with small corrections from mixing with excitation states. The mixing with the excitation states corresponds to virtual processes and is generically independent of initial time. To see this point, one may adapt the formalism presented later in this subsection to many body problems by replacing the Hamiltonian and wavefunctions by the many body ones. The approximate solution to the Schrodinger equation is independent of initial time (up to an overall phase factor). Therefore, we expect the expectation value of observables at any time is independent of the choice of initial time. 
	
	For noninteracting system, it would be desirable to express the expection value of observables in terms of single particle wavefunctions. Generically, the single particle time dependent Schrodinger equation in quasi-1D system is very sensitive to the initial condition. However, as we argued above, the expectation of observables is independent of initial time. Therefore, the strategy below is to pick one electron in the filled states and study its time evolution with a convenient choice of initial time. Then, evaluate the expectation value of observables and lastly sum over all the electrons in the filled states. 
	
	More specifically, we consider the single-particle time-dependent Schrodinger equation for the quasi-1D setup:
	\begin{equation}
		i\partial_t\Phi_{n_0,\kappa_y,k_x}(y,t)=\hat{H}(k_x,t)\Phi_{n_0,\kappa_y,k_x}(y,t)
	\end{equation}
	where the quasi-1D Bloch Hamiltonian $\hat{H}(k_x,t)$ depends on the momentum $k_x$ in the infinite direction. The instantaneous eigenvalues and eigenwavefunctions for the time-dependent Hamiltonian $\hat{H}(k_x,t)$ is denoted as $E_{n/m,\kappa_y,k_x}(t)$ and $U_{n/m,\kappa_y,k_x}(y,t)$. Notice that $U_{n/m,\kappa_y,k_x}(y,t)$ should be understood as the periodic part of the instantaneous Bloch wavefunction. In addition, the time-dependent wavefunction $\Phi_{n,\kappa_y,k_x}(y,t)$ is a function of the transverse position $y$ and time $t$. The wavefunction is labeled by the momentum $k_x$ in $x$-direction as well as the quantum numbers $(n,\kappa_y)$. This way of labeling is associated with the particular form of solution we are seeking in the next paragraph. For brevity, we will denote the quantum numbers as $\tilde{n}/\tilde{m}:=(n/m,\kappa_y)$ with $\tilde{n}=1,2,\cdots,\tilde{N}$ and $\tilde{m}=\tilde{N}+1,\tilde{N}+2,\cdots,\tilde{N}+\tilde{M}$.

	The rest of the derivation is very similar to that of Hall conductance in Ref.~[\onlinecite{shen2012topological}]. Below, we outline a few key steps. In particular, we would like to seek solutions of the following form ($\tilde{n}/\tilde{m}:=(n/m,\kappa_y)$):
	\begin{equation}
		\begin{split}
			\Phi_{\tilde{n}_0,k_x}(y,t)&=\sum_{\tilde{n}^{\prime}} a_{\tilde{n}^{\prime},k_x}(t)e^{-i\int_{t_0}^{t}dt^{\prime}E_{\tilde{n}^{\prime},k_x}(t^{\prime})}U_{\tilde{n}^{\prime},k_x}(y,t)\\
			&+\sum_{\tilde{m}^{\prime}} a_{\tilde{m}^{\prime},k_x}(t)e^{-i\int_{t_0}^{t}dt^{\prime}E_{\tilde{m}^{\prime},k_x}(t^{\prime})}U_{\tilde{m}^{\prime},k_x}(y,t)
		\end{split}
	\label{Eq:Wavefunction_Assume}
	\end{equation}
	with $a_{\tilde{n}_0,k_x}(t)\sim \mathcal{O}(1)$ on the order of one while all other coefficients being small $a_{\tilde{n}/\tilde{m},k_x}(t)\ll 1$ for all $\tilde{n},\tilde{m}\neq\tilde{n}_0$. This form of wavefunction is reasonable because the instantaneous eigenwavefunctions $U_{\tilde{n}/\tilde{m},k_x}(y,t)$ form a complete basis for the Hilbert space of single particle states with momentum $k_x$. The assumption for the coefficients is also possible with a properly chosen initial time $t_0$. 
	
	
	With the assumed form of the wavefunction of Eq.~(\ref{Eq:Wavefunction_Assume}), we may rewrite the time dependent Schrodinger equation in term of the coefficients of $a_{\tilde{n}/\tilde{m},k_x}(t)$. 
	
	For higher energy levels, the coefficients satisfy the following equation:
	\begin{equation}
	\begin{split}
			i\dot{a}_{\tilde{m},k_x}(t)=&-\sum_{\tilde{n}^{\prime}} e^{i\int_{t_0}^{t}dt^{\prime}\left[E_{\tilde{m},k_x}(t^{\prime})-E_{\tilde{n}^{\prime},k_x}(t^{\prime})\right]}\\
			&\times \left[\sum_{\tilde{y}}U^{\dagger}_{\tilde{m},k_x}(\tilde{y},t)\ i\partial_tU_{\tilde{n}^{\prime},k_x}(\tilde{y},t)\right]\ a_{\tilde{n}^{\prime},k_x}(t)
	\end{split}
	\label{Eq:Eq_Coeff_HigherLevels}
	\end{equation}
	Notice that on the right hand side, only the coefficients associated with the lower energy levels labeled by $\tilde{n}$ is summed over. The summation in the second line is over the transverse position, $\tilde{y}$. To solve Eq.~(\ref{Eq:Eq_Coeff_HigherLevels}), one should notice that the phase factor on the first line of the righthand side is the fast dynamics. This is due to the presumed finite energy gap between lower and higher energy levels and the slow evolution of the system. Therefore, the coefficients can be solved approximately as:
	\begin{equation}
		\begin{split}
			a_{\tilde{m},k_x}(t)\approx&\sum_{\tilde{n}^{\prime}}e^{i\int_{t_0}^{t}dt^{\prime}\left[E_{\tilde{m},k_x}(t^{\prime})-E_{\tilde{n}^{\prime},k_x}(t^{\prime})\right]}\\
			&\times \frac{\sum_{\tilde{y}}U^{\dagger}_{\tilde{m},k_x}(\tilde{y},t)\ i\partial_tU_{\tilde{n}^{\prime},k_x}(\tilde{y},t)}{E_{\tilde{m},k_x}(t)-E_{\tilde{n}^{\prime},k_x}(t)}\ a_{\tilde{n}^{\prime},k_x}(t)
		\end{split}
	\label{Eq:GenericSolution_HigherBandsCoeff}
	\end{equation}
	
	Next, we discuss the coefficients associated with the lower energy levels. The equation for the coefficients is similar to Eq.~(\ref{Eq:Eq_Coeff_HigherLevels}):
	\begin{equation}
		\begin{split}
			i\dot{a}_{\tilde{n},k_x}(t)=-&\sum_{\tilde{n}^{\prime}}e^{i\int_{t_0}^{t}dt^{\prime}\left[E_{\tilde{n},k_x}(t^{\prime})-E_{\tilde{n}^{\prime},k_x}(t^{\prime})\right]}\\
			\times &\left[\sum_{\tilde{y}}U^{\dagger}_{\tilde{n}^{\prime},k_x}(\tilde{y},t)\ i\partial_tU_{\tilde{n},k_x}(\tilde{y},t)\right]\ a_{\tilde{n}^{\prime},k_x}(t)
		\end{split}
		\label{Eq:Eq_Coeff_LowerLevels}
	\end{equation}
	with the initial condition of $a_{\tilde{n},k_x}(t_0)=\delta_{\tilde{n},\tilde{n}_0}$. The spectrum of the lower subbands is very dense for wide samples. Thus, all the factors on the right hand side of Eq.~(\ref{Eq:Eq_Coeff_LowerLevels}) can be equally slow. Generically, we need to solve Eq.~(\ref{Eq:Eq_Coeff_LowerLevels}) exactly. However, we may pick the initial time such that the mixing between lower energy levels is negligible. This is possible if the following condition is met:
	\begin{equation}
		\resizebox{0.45\textwidth}{!}{$\left|\sum_{\tilde{y}}U^{\dagger}_{\tilde{n}_0,k_x}(\tilde{y},t)\partial_tU_{\tilde{n},k_x}(\tilde{y},t)\right|(t-t_0)
			\ll\left|\frac{\sum_{\tilde{y}}U^{\dagger}_{\tilde{m},k_x}(\tilde{y},t)i\partial_tU_{\tilde{n}^{\prime},k_x}(\tilde{y},t)}{E_{\tilde{m},k_x}(t)-E_{\tilde{n},k_x}(t)}\right|$}
	\end{equation}
	Following the condition above, Eq.~(\ref{Eq:Eq_Coeff_LowerLevels}) can be solved as:
	\begin{equation}
		a_{\tilde{n},k_x}(t)\approx\delta_{\tilde{n},\tilde{n}_0}
	\end{equation}
	
	Thus, the desired single particle wavefunction is:
	\begin{equation}
	\resizebox{0.5\textwidth}{!}{
		$\begin{split}
		&\Phi_{\tilde{n}_0,k_x}(y,t)=e^{-i\int_{t_0}^{t}dt^{\prime}E_{\tilde{n}_0,k_x}(t^{\prime})}\\
		&\times\left[U_{\tilde{n}_0,k_x}(y,t)+\sum_{\tilde{m}}U_{\tilde{m},k_x}(y,t)\frac{\sum_{\tilde{y}}U^{\dagger}_{\tilde{m},k_x}(\tilde{y},t)\ i\partial_tU_{\tilde{n}_0,k_x}(\tilde{y},t)}{E_{\tilde{m},k_x}(t)-E_{\tilde{n}_0,k_x}(t)}\right]
		\end{split}$
	}
	\label{Eq:Wavefunction}
	\end{equation}
	Notice that the initial time $t_0$ only enters through the overall phase factor and thus does not enter the expectation value of observables.

	The expectation value of an observable $\hat{\mathcal{O}}$ at time $t$ is then given by:
	\begin{equation}
		\langle\hat{\mathcal{O}}\rangle(y,t)=\sum_{\tilde{n}_0}\int\frac{dk_x}{2\pi}\ \Phi^{\dagger}_{\tilde{n}_0,k_x}(y,t)\hat{\mathcal{O}}\Phi_{\tilde{n}_0,k_x}(y,t)
		\label{Eq:Observable}
	\end{equation}
	Here, the summation over all the electrons in the filled states is dictated by the summation over the quantum number $\tilde{n}_0$ and the integration over $k_x$. For observables without momentum derivatives, the expectation value above is indeed independent of the choice of initial time. In Appendix.~\ref{Sec:Observables}, we prove that this is the case by analyzing the solution with a general initial time. In the rest of this section, we will take the observable to be the current density operator. We will provide furthur simplifications and discuss the properties of the current density.

	\subsection{Adiabatic Current}
	
	Following the discussion in the previous subsection, it's straightforward to write down the current between unit cells $(x,y)$ and $(x+1,y)$ by inserting the current operator into Eq.~(\ref{Eq:Observable}). Remember that this is for a quasi-1D sample which is infinite in $x$-direction. The resultant expression for the adiabtic current is the main result of this article. Moreover, the expectation value of the current can be furthur simplified to be:
	\begin{equation}
	\begin{split}
			j_x(y,t)=&\sum_{n,m,\kappa_y}\int\frac{dk_x}{2\pi}\ 2\text{Im}\left[U^{\dagger}_{n,\kappa_y,k_x}(y,t)\ \hat{j}_x\ U_{m,\kappa_y,k_x}(y,t)\right.\\
			&\left.\times\frac{\sum_{\tilde{y}}U^{\dagger}_{m,\kappa_y,k_x}(\tilde{y},t)\ \partial_tU_{n,\kappa_y,k_x}(\tilde{y},t)}{E_{n,\kappa_y,k_x}(t)-E_{m,\kappa_y,k_x}(t)}\right]
	\end{split}
	\label{Eq:MainResult}
	\end{equation}
	where $\hat{j}_x$ is the current operator in momentum space (see Eq.~(\ref{Eq:CurrentOperatorInX})). Here, we restore the full notation for the quantum numbers $\tilde{n}/\tilde{m}=(n/m,\kappa_y)$. The main simplification from Eq.~(\ref{Eq:Observable}) to Eq.~(\ref{Eq:MainResult}) is the following: (i) the term without time derivative is neglected; (ii) the same transverse quantum number $\kappa_y$ appears for both lower and higher energy states. For practical purpose, e.g., a wide or semi-infinite sample, the summation over the transverse quantum number can be replaced by integeration:
	\begin{equation}
		\sum_{\kappa_y}\int\frac{dk_x}{2\pi}\approx (W+1)\int_{\text{BZ}}\frac{dk_xd\kappa_y}{(2\pi)^2}+\sum_{\text{ES}}\int\frac{dk_x}{2\pi}.
		\label{Eq:SumToIntKappaY}
	\end{equation}
	The first term on the right hand side corresponds to the summation over the extended states, where $(W+1)$ is the width of the sample and the integration limit is essentially the 2D Brillouin zone (BZ), even though $\kappa_y$ is the transverse quantum number rather than the momentum in $y$-direction. The second term corresponds to the summation over the edge states (ES) and the integration of $k_x$ is over the quasi-1D BZ.
	
	First of all, in Eq.~(\ref{Eq:Observable}), the term without time derivative:
	\begin{equation}
		j^{\text{eq}}_x(y,t)=\sum_{n,\kappa_y}\int\frac{dk_x}{2\pi}\ U^{\dagger}_{n,\kappa_y,k_x}(y,t)\ \hat{j}_x\ U_{n,\kappa_y,k_x}(y,t)
		\label{Eq:ZerothOrderTerm}
	\end{equation}
	is not included in the main result Eq.~(\ref{Eq:MainResult}). This term corresponds to the equilibrium current at each instant of time. One example is the itinerant boundary current originating from the bulk orbital magnetic moment\cite{vanderbilt2018berry,RevModPhys.82.1959,PhysRevB.100.054408,PhysRevLett.95.137204,PhysRevLett.95.137205}. This term should also vanish deep in the bulk. More importantly, the equilibrium current (if any) does not lead to charge accumulation. Thus, Eq.~(\ref{Eq:ZerothOrderTerm}) is not responsible for any change of the corner charge and is not included in the main result, Eq.~(\ref{Eq:MainResult}). The main result, Eq.~(\ref{Eq:MainResult}), contains first order time derivative. Thus, the adiabatic current is essentially the linear response to the time derivative of the adiabatic parameter.
	
	Furthur simplication to Eq.~(\ref{Eq:MainResult}) relies on the algebraic structure of the quasi-1D Bloch wavefunctions, which is also essential for the discussion of localization properties of the adiabatic current. For demonstration purpose, we consider the simplist situation:
	\begin{equation}
		\begin{split}
			U_{n/m,\kappa_y,k_x}(y,t)=&\alpha_1e^{i\kappa_yy}u^{\text{2D}}_{n/m}(k_x,\kappa_y,t)\\
			+&\alpha_2e^{-i\kappa_yy}u^{\text{2D}}_{n/m}(k_x,-\kappa_y,t)
		\end{split}
	\label{Eq:Quasi1DWF}
	\end{equation}
	where $u^{\text{2D}}_{n/m}(k_x,\kappa_y,t)$ is the periodic part of the 2D bulk Bloch wavefunction with momentum $\boldsymbol{k}=(k_x,\kappa_y)$; $\alpha_{1,2}$ are the combination coefficients, determined from boundary condition. The presence of the plane wave factor in Eq.~(\ref{Eq:Quasi1DWF}) has profound implications. Below, we comment on the simplification from Eq.~(\ref{Eq:Wavefunction}) and Eq.~(\ref{Eq:Observable}) to Eq.~(\ref{Eq:MainResult}). In the next subsection, we show that the adiabatic current is confined to the boundary.
	
	Notice that in the expression of the wavefunction, Eq.~(\ref{Eq:Wavefunction}), or the adiabatic current, Eq.~(\ref{Eq:MainResult}), there is a factor of the form $\sum_{\tilde{y}}U^{\dagger}_{m,\kappa^{\prime}_y,k_x}(\tilde{y},t)\ i\partial_tU_{n,\kappa_y,k_x}(\tilde{y},t)$. Technically, this term involves summation of the form $\sum_{\tilde{y}}\text{Exp}\left[i(\pm\kappa_y\pm\kappa^{\prime}_y)\tilde{y}\right]$, which vanishes for wide samples unless $\kappa_y=\kappa^{\prime}_y$. Therefore, in Eq.~(\ref{Eq:MainResult}), the same transverse quantum number $\kappa_y$ appears in the subscript of the wavefunctions of both lower and higher energy states.
	
	The total adiabatic current along the edge (bottom edge in Fig.~\ref{Fig:SystemDemo}-\ref{Fig:SystemDemoQuasi1D} as an example) is given by:
	\begin{equation}
		j_x(t)=\sum_{y\in\text{bottom edge}}j_x(y,t)
	\end{equation}
	The adiabatic current $j_y$ along the vertical edges of Fig.~\ref{Fig:SystemDemo} can be obtained in exactly the same way by studying quasi-1D samples along $y$-direction. By charge conservation, the change of the bottom left corner charge is given by:
	\begin{equation}
		-\frac{dQ}{dt}=j_x+j_y
	\end{equation}
	where $j_{x(y)}$ is the adiabatic current along the bottom (left) edge.

	\subsection{Localization Property}
	
	In this section, we will show that the adiabatic current is localized at the edges of the sample. To see this point, we still need to employ the specific form of the quasi-1D wavefunction in Eq.~(\ref{Eq:Quasi1DWF}). 
	
	In principle, the adiabatic current, Eq.~(\ref{Eq:MainResult}), contains a position-independent and position-dependent part. Here, we neglect the position-independent part. This is because any nonzero position independent adiabatic current corresponds to the presence of macroscopic bulk current. Namely, charge will be pumped from one side of the sample to the other side. As a result, a dipole moment is generated. This situation violates the presumed inversion symmetry. Thus, below we assume the $y$-independent part in Eq.~(\ref{Eq:MainResult}) vanishes identically and focus on the $y$-dependence.
	
	As discussed in the previous subsection, the adiabatic current in Eq.~(\ref{Eq:MainResult}) can be separated into that carried by the edge states and by extended states. The current carried by the edge states are localized at the edges by the nature of the current carrying states. More interestingly is the current carried by the extended states. Below, we demonstrate that the current carried by the extended states is also localized at the edges.
	
	Direct calculation shows that the adiabatic current associated with the extended states is of the form:
	\begin{equation}
		j^{\text{ex}}_{x}(y,t)=\int\frac{d\kappa_y}{2\pi}\left[f(\kappa_y)\cos(2\kappa_yy)+g(\kappa_y)\sin(2\kappa_yy)\right]
	\end{equation}
	In general, $y$ may be undertood as the distance from the edge. It's clear that $j_{x}(y,t)$ is the $2y$-th Fourier components of two smooth functions $f(\kappa_y)$ and $g(\kappa_y)$. It's mathematical theorem that for $k$-th differentiable periodic functions, the $n$-th Fourier component is asymptotically smaller than $n^{-k}$. Thus, for smooth functions $f(\kappa_y)$ and $g(\kappa_y)$, the adiabatic current asymptotically decays faster than any power law:
	\begin{equation}
		j_x(y,t)<(2y)^{-k}\ \text{for }\forall\ k\in \mathbb{Z}.
	\end{equation}
	Even though the extended states carries current, the charge is only pumped through the edges.

	\subsection{Gauge Invariance}

	To complete the discussion, we should mention that the adiabatic current in Eq.~(\ref{Eq:MainResult}) is gauge independent. To see this point, we rewrite Eq.~(\ref{Eq:MainResult}) to be:
	\begin{equation}
		\begin{split}
			j_x(y,t)=&\sum_{n,m,\kappa_y}\int\frac{dk_x}{2\pi}\ 2\text{Im}\left[U^{\dagger}_{n,\kappa_y,k_x}(y,t)\ \hat{j}_x\ U_{m,\kappa_y,k_x}(y,t)\right.\\
			&\left.\times\frac{\sum_{\tilde{y}}U^{\dagger}_{m,\kappa_y,k_x}(\tilde{y},t)\ \partial_t\hat{H}(k_x,t)\ U_{n,\kappa_y,k_x}(\tilde{y},t)}{\left[E_{n,\kappa_y,k_x}(t)-E_{m,\kappa_y,k_x}(t)\right]^2}\right]
		\end{split}
	\label{Eq:MainResultVariant}
	\end{equation}
	where the following relation is used:
	\begin{equation}
		\begin{split}
			&\sum_{\tilde{y}}U^{\dagger}_{m,\kappa_y,k_x}(\tilde{y},t)\ i\partial_tU_{n,\kappa_y,k_x}(\tilde{y},t)\\
			=&\frac{\sum_{\tilde{y}}U^{\dagger}_{m,\kappa_y,k_x}(\tilde{y},t)\ i\partial_t\hat{H}(k_x,t)\ U_{n,\kappa_y,k_x}(\tilde{y},t)}{E_{n,\kappa_y,k_x}(t)-E_{m,\kappa_y,k_x}(t)}
		\end{split}
	\end{equation}  
	Apparently Eq.~(\ref{Eq:MainResultVariant}) is $U(1)$ gauge independent.
	
	The other situation is when certain energy level is degenerate with degeneracy $G$. The expression Eq.~(\ref{Eq:MainResultVariant}) is also invariant under SU$(G)$ transformation of the basis for the states at that particular degenerate energy level. This is because the wavefunctions $U_g$ of the degenerate energy level come in combination of $\sum_{g} U_gU_g^{\dagger}$ for short. This is an SU$(G)$ invariant object. The adiabatic current is independent of the basis choice for any degenerate energy levels.
	
	Thus, indeed, the derived current, Eq.~(\ref{Eq:MainResult}), is a gauge-independent quantity and represents a physical observable.

	\section{Benalcazar-Bernevig-Hughes (BBH) Model}
	\label{Sec:BBH_Model}

	In this section, we apply the derived adiabatic current in the previous section to the BBH model. BBH model is the first example of higher order topological insulators (HOTI)\cite{Benalcazar61,PhysRevB.96.245115}. Instead of working in the situations with quantized corner charge, we introduce an on-site sublattice potential that breaks the protecting symmetry of HOTI phase (rotation and reflections). In this situation, the corner charge is not quantized and may change continuously by varying the on-site potential and the hopping parameters. We show that the derived adiabatic current has good agreement with the change of the corner charge. We should mention that the discussion in this section is very similar to that of SSH model\cite{PhysRevB.22.2099} and Thouless charge pump\cite{PhysRevB.27.6083} in one dimension. 
	
	\begin{figure}[tb]
		\centering
		\includegraphics[scale=0.35]{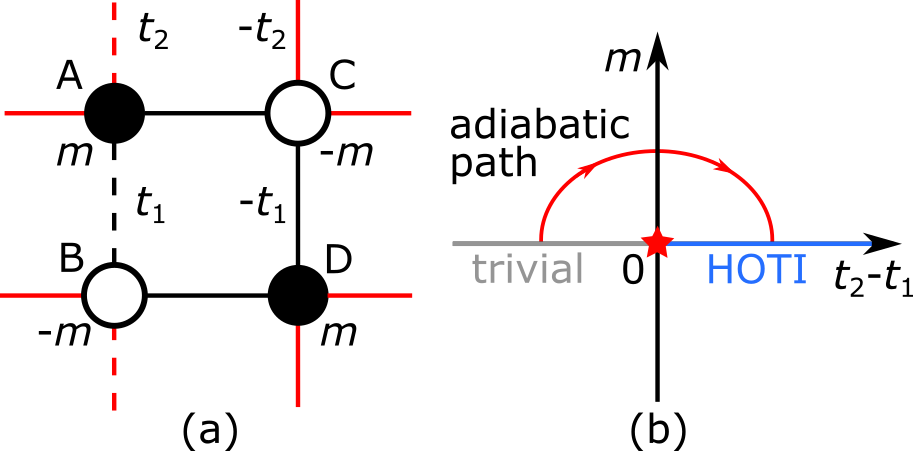}
		\caption{(a) Unit cell of BBH model, which forms square lattice as in Fig.~\ref{Fig:SystemDemo}. Each unit cell contains four sublattice sites, labeled from A to D, with onsite sublattice potential $m_{\text{A,D}}=-m_{\text{B,C}}=m$. The magntitude of the intra (inter) unit cell hopping is $t_{1(2)}$. In addition, the hopping parameter flips sign on the bonds indicated by the dashed lines, due to the presumed threading of $\pi$-flux in each plaquett. (b) Phase diagram of BBH model. Without sublattice potential $m=0$, there is a gap closing point (red star) separating the phase of trivial insulator (gray) and HOTI (blue). With sublattice potential, the trivial and HOTI phase can be adiabatically connected following the adiabatic path (red path). Here, we assumed $t_{1,2}>0$ for clarity.}
		\label{Fig:BBHModel}
	\end{figure}
	
	
	\begin{figure}[tb]
		\centering
		\includegraphics[scale=0.45]{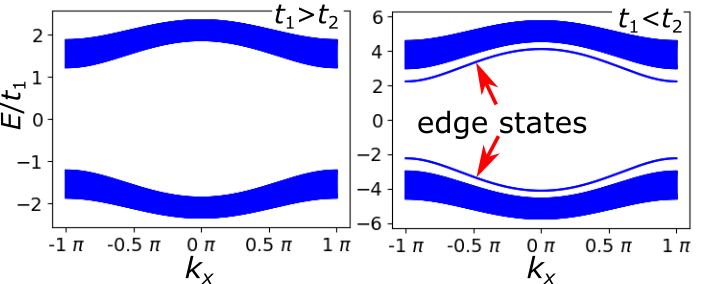}
		\caption{Quasi-1D band structure of BBH model with the width of 40 unit cells at two typical choices of parameters. The BBH model is fully insulating in the quasi-1D geometry. Insulating edge states appear in the case of $t_1<t_2$. The parameters are chosen as $m=t_1=1$ and (left) $t_2=0.5$;(right) $t_2=3$.}
		\label{Fig:BBH1DSpectrum}
	\end{figure}

	The unit cell of BBH model is depicted in Fig.~\ref{Fig:BBHModel}, which further forms a square lattice in Fig.~\ref{Fig:SystemDemo}. Each unit cell contains four sublattice sites labeled from $A$ to $D$, forming a plaquett. The intra unit cell hopping is parameterized as $t_1$, while the inter unit cell hopping as $t_2$. Additionally, each plaquett has $\pi$ magnetic flux. As a result, one of the hopping parameter in each plaquett flips sign, as indicated by the dashed line in Fig.~\ref{Fig:BBHModel}(a).
	
	Without the on-site sublattice potential, BBH support two distinct phases of matter, Fig.~\ref{Fig:BBHModel}(b). When the intra-unit-cell hopping is stronger than the inter-unit-cell one $t_1>t_2$, the situation is termed as trivial due to the fully gapped spectrum. In the opposite situation $t_2>t_1$, the 2D bulk spectrum is fully gapped. However, for a sample of square geometry, four degenerate corner states appear. Thus, this phase is dubbed as higher order topological insulator. The corners hold fractional $\pm\frac{1}{2}$ excess charge. Without the sublattice potential, the trivial and the topological phase cannot be adiabatically connected. A bulk gap closing point at $t_1=t_2$ is unavoidable.
	
	The gap closing point may be avoided by introducing the on-site sublattice potential. In this section, we introduce the following sublattice potential:
	\begin{equation}
		m_{\text{A,D}}=-m_{\text{B,C}}=m
	\end{equation}
	The onsite potential breaks the reflection and the $\mathcal{C}_4$ rotational symmetry of the original BBH model\cite{PhysRevB.96.245115,Benalcazar61}, while it fully respects the inversion symmetry. In this case, the trivial and topological phase of BBH model lose their symmetry protection and thus can be adiabatically connected, as indicated by the red path in Fig.~\ref{Fig:BBHModel}(b). Fig.~\ref{Fig:BBH1DSpectrum} shows the quasi-1D spectra of BBH model at two typical choice of the parameters to be fully gapped. In particular, gapped edge states emerge when the intra unit cell hopping is weaker than the inter unit cell one $t_1<t_2$. Thus, with any choice of parameters, the system is fully gapped and dipole free . The corner charge can be unambigiously defined.
	
	We furthur consider the following adiabatic pumping process:
	\begin{equation}
		\left\{ \begin{split}
			&m=M\sin\theta(t),\ \ \ t_2=t_1-\lambda\cos\theta(t)\\
			&\text{with }\theta(t)=\frac{\pi}{T}t\ \ \ \text{for }0<t<T,
		\end{split}\right.
	\end{equation}
	adiabatically connecting the trivial phase at $t=0$ and the HOTI phase at $t=T$. At each instant of time, the quasi-1D spectrum is fully gapped.

	At $t=0$, the system is in the phase of a trivial insulator. The corner charge vanishes. We took the state at $t=0$ as a reference point to define the excess corner charge in the latter time of the adiabatic pumping process. Numerically, we diagonalized a sample of $40\times40$ unit cells at each instant of time of the pumping process. Each corner is a square of $10\times10$ unit cells. The corner charge is defined as:
	\begin{equation}
			Q_{\text{c}}=\sum_{\epsilon_i(t)<0}\sum_{\boldsymbol{R}\in\text{Corner}}\left[\psi^{\dagger}_i(\boldsymbol{R},t)\psi_i(\boldsymbol{R},t)-\psi^{\dagger}_i(\boldsymbol{R},0)\psi_i(\boldsymbol{R},0)\right]
			\label{Eq:CornerChargeDef}
	\end{equation}
	where $\epsilon_i(t)$ and $\psi_i(\boldsymbol{R},t)$ are the instantaneous energy eigenvalue and eigenwavefunction; the summation is over all the states with negative energy and positions in one of the corner regions. One may equally choose the bulk charge density at each instant of time as reference. The difference should vanish in the thermodynamic limit.
	
	\begin{figure}[t]
		\includegraphics[scale=0.6]{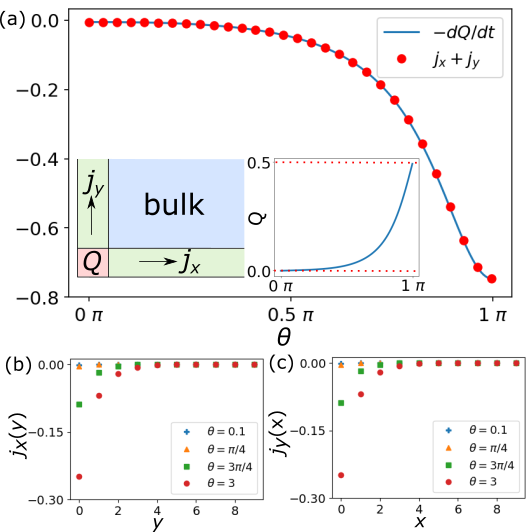}
		\caption{(a) The sum of adiabatic currents along the left and bottom edge, $j_x+j_y$, for the adiabatic pumping process (plotted by the red dots), obtained from Eq.~(\ref{Eq:MainResult}).  This is to be compared with the negative of the changing rate of the left-bottom corner charge, $-dQ/dt$ (plotted by solid blue line), from numerical derivative of the corner charge. The unit for the vertical axis is $\dot{\theta}=\frac{\pi}{T}$. Inset: Left: Schematic separation of the sample into corner (red), edge (green) and bulk (blue) region. Right: Left-bottom corner charge $Q$ during the adiabatic evolution process, obtained from Eq.~(\ref{Eq:CornerChargeDef}). (b) The bottom edge adiabatic current $j_x(y)$  as function of distance $y$ from the bottom boundary. (c) The left edge adiabatic current $j_y(x)$  as function of distance $x$ from the left boundary. The parameters for the adiabatic process are chosen as $t_1=M=1$ and $\lambda=0.5$.}
		\label{Fig:CC_Current}
	\end{figure}
	
	Due to the specific form of the sublattice potential, the electrons flow towards the left-bottom and the right-top corner during the adiabatic current. Indeed, as shown in the inset of Fig.~\ref{Fig:CC_Current}(a), the left-bottom corner charge continously increases from zero at $t=0$ to one half at $t=T$. The changing rate of the corner charge in Fig.~\ref{Fig:CC_Current}(a) is obtained by taking numerical derivative of the corner charge with respect to time (or equivalently the adiabatic parameter $\theta(t)$). 
	
	By charge conservation, the change of the left-bottom corner charge is associated with the adiabatic current flowing along the bottom and the left edges:
	\begin{equation}
		-\frac{dQ}{dt}=j_x+j_y.
	\end{equation}
	The adiabatic current is obtained by inserting quasi-1D wavefunctions (see Appendix.~\ref{Sec:DetailsOnTB}) into the main result, Eq.~(\ref{Eq:MainResult}) and then summing over the edge region of $10$ unit cell wide. The adiabatic current is proportional to the time derivative of the adiabatic parameter, $\dot{\theta}=\pi/T$, which is factored out in the Fig.~\ref{Fig:CC_Current}. Fig.~\ref{Fig:CC_Current}(b-c) plots the distribution of the adiabatic current in the edge region. It shows that the adiabatic current along the edges are vanishingly small just a few unit cells away from the system boundary. The similarity of Fig.~\ref{Fig:CC_Current}(b) and (c) is expected to be a special property of BBH model\cite{Benalcazar61,PhysRevB.96.245115}.
	
	Most impoartantly, Fig.~\ref{Fig:CC_Current}(a) shows a good agreement between the numerically obtained changing rate of the corner charge and the proposed equation of the adiabatic current, Eq.~(\ref{Eq:MainResult}). Thus, we conclude that the main result of the adiabatic current, Eq.~(\ref{Eq:MainResult}), can accurately predict the change of the corner charge. It applies to systems with gapped bulk and quasi-1D spectrum. Furthurmore, an integration over the adiabatic current could accurately predict the charge accumulation at the corners.

	\section{Conclusion and Discussion}
	\label{Sec:Conclusion}

	To conclude, in this article, we aimed at predicting the change of corner charge in a two dimensional insulator with inversion symmetry. We studied the adiabatic evolution of the system and derived an analytical expression for the adiabatic current distribution in quasi-1D geometry infinite in $x$-direction:
	\widetext
	\begin{equation}
			j_x(y,t)=\sum_{n,m,\kappa_y}\int\frac{dk_x}{2\pi}\ 2\text{Im}\left[U^{\dagger}_{n,\kappa_y,k_x}(y,t)\ \hat{j}_x\ U_{m,\kappa_y,k_x}(y,t)
		\frac{\sum_{\tilde{y}}U^{\dagger}_{m,\kappa_y,k_x}(\tilde{y},t)\ \partial_tU_{n,\kappa_y,k_x}(\tilde{y},t)}{E_{n,\kappa_y,k_x}(t)-E_{m,\kappa_y,k_x}(t)}\right]
		\label{Eq:MainResult1}
	\end{equation}
	\twocolumngrid
	The wavfunctions $U_{n/m,\kappa_y,k_x}(y,t)$ are the periodic part of the quasi-1D Bloch wavefunctions, with momentum $k_x$ and the transverse quantum number $\kappa_y$ as well as other quantum numbers $n/m$. $\hat{j}_x$ is the current operator along horizontal bonds in momentum space. For practical purpose, one may replace the summation over the transverse quantum number $\kappa_y$ by an integration according to Eq.~(\ref{Eq:SumToIntKappaY}). Notice that the adiabatic current is essentially the linear response to the time derivative of the adiabatic parameter, due to the presence of the first order time derivative. For numerical computation, one may employ Eq.~(\ref{Eq:MainResultVariant}) as an alternative version of the main result, to avoid taking numerical derivative on the wavefunctions.

	Summation of the transverse position $y$ over a few unit cell layers in the edge region gives the net adiabatic current along the edge in $x$-direction. This result applies to systems, whose quasi-1D spectrum is fully gapped. The main result, Eq.~(\ref{Eq:MainResult}) or Eq.~(\ref{Eq:MainResult1}), could be straightforwardly adapted to other quasi-1D orientations. Thus the current formulation can be applied to study the corner charge of various sample geometries with sharp edges, such as polygons.  Indeed, we showed that the sum of the adiabatic current along the edges cutting an corner can accurately predict the change of the corner charge in the case of BBH model.

	This work provides a different perspective on the prediction of corner charge. In particular, we showed explicitly that the adiabatic current is localized at the edges of the sample. The definition of the adiabatic current is gauge invariant. Certain technical challenges, such as picking consistent gauge and constructing hybrid Wannier functions, are avoided\cite{PhysRevB.96.245115,PhysRevB.103.035147}.
	
	We emphasize that the adiabaticity for a quasi-1D system really comes from the gap in the many body spectrum. Then, certain simplifications can be made for noninteracting system to express the expectation values of observables in terms of single particle wavefunctions. It would be interesting to develop a purely many body formulation for the adiabatic current. We leave the many body generalization as a future work, which should be subject to careful checking. Major challenge is that with a limited numerically accessible system size, the computation of corner charge may require certain convolution and smearing scheme to limit the finite size corrections\cite{PhysRevB.103.035147,vanderbilt2018berry}. 
	
	It's important to notice that there is no obvious separation of the adiabatic current into ``purely bulk'' or ``purely edge'' contributions (aside from the current by the edge states). This is because the wavefunctions in the main result, Eq.~(\ref{Eq:MainResult1}), are the quasi-1D wavefunctions which extends over the whole sample. At the same time, the boundary condition is important. Namely, certain coefficients ($\alpha_i$ in Eq.~(\ref{Eq:Quasi1DWF})) determined by the boundary condition necessarily enter the expression of the adiabatic current. In this article, we assumed perfect crystals, which are commensurate with the choice of unit cell and have perfect boundaries. In certain cases, surface relaxation may take place and change the boundary conditions. The quasi-1D Bloch wavefunctions will be altered accordingly. Thus, the adiabatic current as well as the corner charge will receive corrections from surface relaxation\cite{PhysRevB.92.041102}. This is an evidence suggesting that the corner charge is not a purely bulk or purely edge effect, but rather a mixed one. In addition, it's not obvious that certain part of the adiabatic current is responsible for the change of edge polarization and the other part for the bulk quadrupole moment. Thus, the edge polarization and the bulk quadrupole moment are not independently defined. This is in accordance with the previous work, which shows that any definition of edge polarization or bulk quadrupole moment is gauge dependent\cite{PhysRevB.103.035147}. Indeed, the physical observable is the corner charge, rather than the edge polarization or the bulk quadrupole moment.

	Lastly, during the preparation of this manuscript, we noticed the preprint Ref.~[\onlinecite{may2021crystalline}]. In this work, the authors proposed a field theoretic approach to predict the corner charge or bound charge at the defect for $\mathcal{C}_4$ rotational symmetric insulators. Ref.~[\onlinecite{may2021crystalline}] also shows that with interaction, the corner charge fractionalizes. It would be interesting to see if their formulation can be adapted to a generic situation or if there is any deep connection with the current work.
	
	\section{Acknowlegement}
	
	We are grateful to A. A. Burkov for discussion and A. Kamenev and Hanteng Wang for the suggestions during the preparation of the manuscript. This work is supported by Natural Sciences and Engineering Research Council (NSERC) of Canada.

	\appendix
	
	
	\section{Observables Independent of Initial Time}
	\label{Sec:Observables}
	
	In this section, we prove that the expectation value of any observable is indeed independent of initial time. Still, we focus on the noninteracting systems. The general consideration is the same as in Sec.~\ref{Sec:AdiaEvo_ExpValObser}.
	
	Here, we focus on Eq.~(\ref{Eq:Eq_Coeff_LowerLevels}), which we rewrite compactly as follows:
	\begin{equation}
		i\dot{a}_{\tilde{n},k_x}(t)=\sum_{\tilde{n}^{\prime}}A^{k_x}_{\tilde{n}\tilde{n}^{\prime}}(t)\ a_{\tilde{n}^{\prime},k_x}(t)
		\label{Eq:Coeff_lowerbands_variant}
	\end{equation}
	with the coefficients on the righthand side forming an $\tilde{N}\times\tilde{N}$ Hermitian matrix, whose elements are:
	\begin{equation}
		\begin{split}
			A^{k_x}_{\tilde{n}\tilde{n}^{\prime}}(t)=-&e^{i\int_{t_0}^{t}dt^{\prime}\left[E_{\tilde{n},k_x}(t^{\prime})-E_{\tilde{n}^{\prime},k_x}(t^{\prime})\right]}\\
			\times &\sum_{\tilde{y}}U^{\dagger}_{\tilde{n}^{\prime},k_x}(\tilde{y},t)\ i\partial_tU_{\tilde{n},k_x}(\tilde{y},t)
		\end{split}
	\end{equation}
	Generically, the solution of Eq.~(\ref{Eq:Coeff_lowerbands_variant}) can be formally written as:
	\begin{equation}
		a_{\tilde{n},k_x}(t)=\sum_{\tilde{n}^{\prime}}\Lambda^{tt_0,k_x}_{\tilde{n}\tilde{n}^{\prime}}a_{\tilde{n}^{\prime},k_x}(t_0)
		\label{Eq:GenericSolution_LowerBandsCoeff}
	\end{equation}
	where $\Lambda^{tt_0,k_x}$ is an $\tilde{N}\times\tilde{N}$ unitary matrix, formally given by:
	\begin{equation}
		\Lambda^{tt_0,k_x}_{\tilde{n}\tilde{n}^{\prime}}=\left[\mathcal{T}e^{-i\int_{t_0}^{t}A^{k_x}(t^{\prime})dt^{\prime}}\right]_{\tilde{n}\tilde{n}^{\prime}}
	\end{equation}
	with $\mathcal{T}$ being time ordering operator.
	
	The solution for the coefficient $a_{\tilde{m},k_x}(t)$ in Eq.~(\ref{Eq:GenericSolution_HigherBandsCoeff}) remains a good approximation for slowly evolving Hamiltonian. Now we consider the wavefunction of an electron initially at the state $\tilde{n}_0$. With the initial condition $a_{\tilde{n},k_x}(t_0)=\delta_{\tilde{n},\tilde{n}_0}$, the wavefunction at time $t$ is given by:
	\begin{equation}
		\resizebox{0.5\textwidth}{!}{
			$\begin{split}
				&\Psi_{\tilde{n}_0,k_x}(y,t)=\sum_{\tilde{n}}\Lambda^{tt_0,k_x}_{\tilde{n}\tilde{n}_0}\ \ e^{-i\int_{t_0}^{t}dt^{\prime}E_{\tilde{n},k_x}(t^{\prime})}\\
				&\times\left[U_{\tilde{n},k_x}(y,t)+\sum_{\tilde{m}}U_{\tilde{m},k_x}(y,t)\frac{\sum_{\tilde{y}}U^{\dagger}_{\tilde{m},k_x}(\tilde{y},t)\ i\partial_tU_{\tilde{n},k_x}(\tilde{y},t)}{E_{\tilde{m},k_x}(t)-E_{\tilde{n},k_x}(t)}\right]
			\end{split}$
		}
	\end{equation}
	Notice that the wavefunction can be rewritten as:
	\begin{equation}
		\Psi_{\tilde{n}_0,k_x}(y,t)=\sum_{\tilde{n}}\Lambda^{tt_0,k_x}_{\tilde{n}\tilde{n}_0}\Phi_{\tilde{n},k_x}(y,t)
		\label{Eq:Generic_Wavefunc}
	\end{equation}
	where $\Phi_{\tilde{n},k_x}(y,t)$ is given by Eq.~(\ref{Eq:Wavefunction}). Indeed, the single particle wavefunction is very sensitive to the initial time. However, below we show that the expectation value of any observables is independent of the initial time, after summing over all the elctrons in the filled states.
	
	In particular, for any observable $\mathcal{O}$ without momentum derivatives, we will show the following relation:
	\begin{equation}
		\begin{split}
			&\sum_{\tilde{n}_0}\int\frac{dk_x}{2\pi}\Psi^{\dagger}_{\tilde{n}_0,k_x}(y,t)\hat{\mathcal{O}}\Psi_{\tilde{n}_0,k_x}(y,t)\\
			=&\sum_{\tilde{n}_0}\int\frac{dk_x}{2\pi}\Phi^{\dagger}_{\tilde{n}_0,k_x}(y,t)\hat{\mathcal{O}}\Phi_{\tilde{n}_0,k_x}(y,t)
		\end{split}
	\label{Eq:IndepenetofTime}
	\end{equation}
	where the summation is over the $\tilde{N}$ electrons initially in the filled states labeled as $\tilde{n}_0$. To see the relation above, first insert Eq.~(\ref{Eq:Generic_Wavefunc}) into the first line of Eq.~(\ref{Eq:IndepenetofTime}):
	\begin{equation}
		\begin{split}
			&\sum_{\tilde{n}_0}\int\frac{dk_x}{2\pi}\Psi^{\dagger}_{\tilde{n}_0,k_x}(y,t)\hat{\mathcal{O}}\Psi_{\tilde{n}_0,k_x}(y,t)\\
			=&\sum_{\tilde{n}_0}\int\frac{dk_x}{2\pi}\sum_{\tilde{n}_1,\tilde{n}_2}\left[\Lambda^{tt_0,k_x}_{\tilde{n}_1\tilde{n}_0}\right]^*\Lambda^{tt_0,k_x}_{\tilde{n}_2\tilde{n}_0}\Phi^{\dagger}_{\tilde{n}_1,k_x}(y,t)\hat{\mathcal{O}}\Phi_{\tilde{n}_2,k_x}(y,t)
		\end{split}
	\end{equation}
	Then, we need to employ the unitarity of the matrix $\Lambda^{tt_0,k_x}$:
	\begin{equation}
		\sum_{\tilde{n}_0}\left[\Lambda^{tt_0,k_x}_{\tilde{n}_1\tilde{n}_0}\right]^*\Lambda^{tt_0,k_x}_{\tilde{n}_2\tilde{n}_0}=\delta_{\tilde{n}_1\tilde{n}_2}.
	\end{equation}
	With the relation above, we derive the relation Eq.~(\ref{Eq:IndepenetofTime}). This relation suggests that the expectation value of any observables is independent of the initial time. Indeed,as in Eq.~(\ref{Eq:Wavefunction}), the initial time only enters as an overall phase factor in the wavefunction $\Phi_{\tilde{n},k_x}(y,t)$ in the second line of Eq.~(\ref{Eq:IndepenetofTime}).
	
	To this end, we have proved that the expectation value of any observables is independent of the choice of initial time for an insulator undegoing adiabatic evolution.

	\section{Details on the Tight Binding Model and BBH Model}
	\label{Sec:DetailsOnTB}

	In this section, we provide the necessary detail for calculating the adiabatic current using Eq.~(\ref{Eq:MainResult}). We first introduce the generic tight binding description of the square lattice shown in Fig.~\ref{Fig:SystemDemo}. We introduce the spectrum and wavefunctions of BBH model for both 2D bulk and quasi-1D geometry.
	
	Generically, in the bulk of the square lattice shown in Fig.~\ref{Fig:SystemDemo}, noninteracting electrons' dynamics is given by the following Hamiltonian:
	\begin{equation}
		\begin{split}
			&H=\sum_{x,y}\psi^{\dagger}(x,y)H_{\text{os}}\psi(x,y)\\
			&+\sum_{x,y}\left[\psi^{\dagger}(x+1,y)T_x\psi(x,y)+\psi^{\dagger}(x,y+1)T_y\psi(x,y)+\text h.c.\right]
		\end{split}
	\label{Eq:Hamil}
	\end{equation}
	where $H_{\text{os}}$, $T_x$ and $T_y$ are $(N+M)\times(N+M)$ matrices, describing intra-unit-cell coupling and the hoppings along the horizontal bonds and vertical bonds respectively; $\psi(x,y)$ is an $(N+M)$ component spinor.
	
	It's standard quantum mechanical calculation to show that the current flowing between the unit cells $(x,y)$ and $(x+1,y)$ is given by:
	\begin{equation}
		j_x=-i\left[\psi^{\dagger}(x+1,y)T_x\psi(x,y)-\text{h.c.}\right].
	\end{equation}
	Equivalently in momentum space, the current operator along horizontal bonds between unit cells is given by:
	\begin{equation}
		\hat{j}_x=-i\left(T_xe^{-ik_x}-T^{\dagger}_xe^{ik_x}\right),
		\label{Eq:CurrentOperatorInX}
	\end{equation}
	which is to be inserted into Eq.~(\ref{Eq:MainResult}).
	
	Similarly, the current between the unit cells $(x,y)$ and $(x, y+1)$ is given by:
	\begin{equation}
		j_y=-i\left[\psi^{\dagger}(x,y+1)T_y\psi(x,y)-\text{h.c.}\right].
	\end{equation}
	In momentum space, the current operator along the vertical bonds between unit cells is given by:
	\begin{equation}
		\hat{j}_y=-i\left(T_ye^{-ik_y}-T^{\dagger}_ye^{ik_y}\right).
		\label{Eq:CurrentOperatorInY}
	\end{equation}
	
	\subsection{BBH model: Spectrum and Wavefunctions}
	For BBH model, the Hamiltonian is given by Eq.~(\ref{Eq:Hamil}), with
	\begin{eqnarray}
		&H_{\text{os}}=\begin{bmatrix}
			m & t_1 & -t_1 & 0\\
			t_1 & -m & 0 & -t_1\\
			-t_1 & 0 & -m & -t_1\\
			0 & -t_1 & -t_1 & m
		\end{bmatrix},\\
	&T_x=\begin{bmatrix}
		\ 0 & \ 0 & -t_2 & 0\\
		\ 0 & \ 0 & 0 & -t_2\\
		\ 0 & \ 0 & 0 & 0\\
		\ 0 & \ 0 & 0 & 0
	\end{bmatrix},\ T_y=\begin{bmatrix}
		0 & 0 & 0 & 0\\
		t_2 & 0 & 0 & 0\\
		0 & 0 & 0 & 0\\
		0 & 0 & -t_2 & 0
	\end{bmatrix}.
	\end{eqnarray}
	The wavefunction $\psi=(\psi_{\text{A}},\psi_{\text{B}},\psi_{\text{C}},\psi_{\text{D}})^{\text{T}}$ is a four component spinor.
	
	This model has four bands. Two lower bands (as well as the two upper bands) are degenerate. The band structure is given by:
	\begin{equation}
		E_{\pm}(\boldsymbol{k})=\pm\sqrt{m^2+2t_1^2+2t_2^2+2t_1t_2\cos k_x+2t_1t_2\cos k_y}.
	\end{equation}
	There is a finite energy gap between the lower and the upper bands, which vanishes only when $t_1=t_2$ and $m=0$.
	
	
	The wavefunctions takes the form of Bloch wavefunction, $\psi_{i,\boldsymbol{k}}(x,y)=e^{i(k_xx+k_yy)}u_i(\boldsymbol{k})$. For the two lower bands with energy $E_-(\boldsymbol{k})$, the periodic part of the wavefunction is:
	\begin{equation}
		\begin{split}
		&u_1(\boldsymbol{k})=\frac{1}{\sqrt{2E(E+m)}}\begin{bmatrix}
			-t_1-t_2e^{ik_y}\\ m + E\\0 \\ t_1+t_2e^{ik_x}
		\end{bmatrix},\\
	&u_2(\boldsymbol{k})=\frac{1}{\sqrt{2E(E+m)}}\begin{bmatrix}
		t_1+t_2e^{-ik_x}\\0\\m+E\\t_1+t_2e^{-ik_y}
	\end{bmatrix}
		\end{split}
	\end{equation}
	while for the two higher bands with energy $E_+(\boldsymbol{k})$, the periodic part of the wavefunction reads:
	\begin{equation}
		\begin{split}
			&u_3(\boldsymbol{k})=\frac{1}{\sqrt{2E(E+m)}}\begin{bmatrix}
				-m-E\\-t_1-t_2e^{-ik_y}\\ t_1+t_2e^{ik_x}\\0
			\end{bmatrix},\\
			&u_4(\boldsymbol{k})=\frac{1}{\sqrt{2E(E+m)}}\begin{bmatrix}
				0 \\ t_1+t_2e^{-ik_x}\\t_1+t_2e^{ik_y}\\-m-E
			\end{bmatrix}
		\end{split}
	\end{equation}
	where $E=\sqrt{m^2+2t_1^2+2t_2^2+2t_1t_2\cos k_x+2t_1t_2\cos k_y}$.
	
	\widetext
	
	\subsection{Quasi-1D geometry along $x$-direction}

	For the quasi-1D geometry infinite in $x$-direction and finite in $y$-direction, the quasi-momentum $k_x$ remains a good quantum number. $\kappa_y$ denotes transverse quantum number. The quasi-1D Bloch wavefunction now takes the form $\psi_{n/m,\kappa_y,k_x}(x,y)=e^{ik_xx}U_{n/m,\kappa_y,k_x}(y)$. The periodic part of the quasi-1D Bloch wavefunction takes the form of:
	\begin{equation}
				U_{n/m,\kappa_y,k_x}(y)=\alpha_1e^{i\kappa_yy}u_{1/3}(k_x,\kappa_y)+\alpha_2e^{-i\kappa_yy}u_{1/3}(k_x,-\kappa_y)		+\alpha_3e^{i\kappa_yy}u_{2/4}(k_x,\kappa_y)+\alpha_4e^{-i\kappa_yy}u_{2/4}(k_x,-\kappa_y)
				\label{Eq:Q1D_WF}
	\end{equation}
	where $\alpha_i$ with $i=1,2,3,4$ is the coefficient of superposition, to be determined from the boundary condition.

	
	The boundary condition is given by:
	\begin{subequations}
		\begin{align}
		&E_{n/m,\kappa_y}(k_x)U_{n/m,\kappa_y,k_x}(0)=h(k_x)U_{n/m,\kappa_y,k_x}(0)+T_y^{\dagger}U_{n/m,\kappa_y,k_x}(1) \label{Eq:xBC_Bottom}\\
		&E_{n/m,\kappa_y}(k_x)U_{n,\kappa_y,k_x}(W)=h(k_x)U_{n/m,\kappa_y,k_x}(W)+T_yU_{n/m,\kappa_y,k_x}(W-1)
		\label{Eq:xBC_Top}
		\end{align}
	\label{Eq:xBC}
	\end{subequations}
	where $h(k_x)=H_{\text{os}}+T_xe^{-ik_x}+T_x^{\dagger}e^{ik_x}$. The quasi-1D spectrum is given by:
	\begin{equation}
		E_{n/m,\kappa_y}(k_x)=\pm\sqrt{m^2+2t_1^2+2t_2^2+2t_1t_2\cos k_x+2t_1t_2\cos \kappa_y}.
	\end{equation}
	The boundary conditions in Eq.~(\ref{Eq:xBC_Bottom}-\ref{Eq:xBC_Top}) determines the combination coefficients $\alpha_i$ and the possible values of $\kappa_y$ in Eq.~(\ref{Eq:Q1D_WF}).
	
	For the extended states, if we want to study the bottom edge, then the boundary condition Eq.~(\ref{Eq:xBC_Bottom}) suggests that the quasi-1D wavefunctions should take the following form: (i) For the lower filled subbands, the periodic part of the quasi-1D Bloch wavefunctions are:
	\begin{eqnarray}
			&U_{1,\kappa_y,k_x}(y)=\frac{e^{i\kappa_yy}}{\sqrt{2(W+1)}}\frac{t_2+t_1e^{i\kappa_y}}{t_2+t_1e^{-i\kappa_y}}u_1(k_x,\kappa_y)-\frac{e^{-i\kappa_yy}}{\sqrt{2(W+1)}}u_1(k_x,-\kappa_y)\\
			&U_{2,\kappa_y,k_x}(y)=\frac{e^{i\kappa_yy}}{\sqrt{2(W+1)}}e^{2i\kappa_y}u_2(k_x,\kappa_y)-\frac{e^{-i\kappa_yy}}{\sqrt{2(W+1)}}u_2(k_x,-\kappa_y)
	\end{eqnarray}
	(ii) For the upper empty subbands, the periodic part of the quasi-1D Bloch wavefunctions are:
	\begin{eqnarray}
			&U_{3,\kappa_y,k_x}(y)=\frac{e^{i\kappa_yy}}{\sqrt{2(W+1)}}e^{2i\kappa_y}u_3(k_x,\kappa_y)-\frac{e^{-i\kappa_yy}}{\sqrt{2(W+1)}}u_3(k_x,-\kappa_y)\\
			&U_{4,\kappa_y,k_x}(y)=\frac{e^{i\kappa_yy}}{\sqrt{2(W+1)}}\frac{t_2+t_1e^{i\kappa_y}}{t_2+t_1e^{-i\kappa_y}}u_4(k_x,\kappa_y)-\frac{e^{-i\kappa_yy}}{\sqrt{2(W+1)}}u_4(k_x,-\kappa_y)
	\end{eqnarray}
	The boundary condition at the top edge, Eq.~(\ref{Eq:xBC_Top}), furthur gives the quantization condition for the transverse quantum number $\kappa_y$:
	\begin{equation}
		\frac{t_2+t_1e^{-i\kappa_y}}{t_2+t_1e^{i\kappa_y}}=e^{2i\kappa_y(W+1)}.
		\label{Eq:QC}
	\end{equation}
	For a wide sample $W\gg1$, the transverse quantum number $\kappa_y$ are very densely valued, with an average distance of $\pi/(W+1)$. Thus, practically, the summation over the transverse quantum number can be replaced by an integration.
	
	It turns out that the quantization condition, Eq.~(\ref{Eq:QC}), has complex solution when $t_1<t_2$. In this case, the transverse quantum number may take the following value:
	\begin{equation}
		\kappa^c_y=\pi+i\ln\frac{t_2}{t_1}\ \ \ \ \ \text{for}\ (W+1)\ln\frac{t_2}{t_1}\gg1
	\end{equation}
	This solution corresponds to the gapped edge states.
	
	There are two degenerate edge bands with energy:
	\begin{equation}
		E^-_{\text{edge}}(k_x)=-E=-\sqrt{m^2+t_1^2+t_2^2+2t_1t_2\cos k_x}
	\end{equation}
	The corresponding wavefunctions are the following:
	\begin{subequations}
		\begin{align}
			&U^{\text{edge}}_{1,k_x}(y)=\left[1-\left(t_1/t_2\right)^2\right]^{1/2}e^{i\pi y}e^{-y\ln\frac{t_2}{t_1}}\frac{1}{\sqrt{2E(E+m)}}\begin{bmatrix}
			0\\m+E\\0\\t_1+t_2e^{ik_x}
		\end{bmatrix}  \label{Eq:ES_LE_Bot}\\
		&U^{\text{edge}}_{2,k_x}(y)=\left[1-\left(t_1/t_2\right)^2\right]^{1/2}e^{i\pi y}e^{(y-W)\ln\frac{t_2}{t_1}}\frac{1}{\sqrt{2E(E+m)}}\begin{bmatrix}
			t_1+t_2e^{ik_x}\\0\\m+E\\0
		\end{bmatrix}
		\label{Eq:ES_LE_Top}
		\end{align}
	\end{subequations}
	
	In addition, there are two degenerate edge bands with energy:
	\begin{equation}
		E^+_{\text{edge}}(k_x)=E=\sqrt{m^2+t_1^2+t_2^2+2t_1t_2\cos k_x}
	\end{equation}
	The corresponding wavefunctions are:
	\begin{subequations}
		\begin{align}
			&U^{\text{edge}}_{3,k_x}(y)=\left[1-\left(t_1/t_2\right)^2\right]^{1/2}e^{i\pi y}e^{-y\ln\frac{t_2}{t_1}}\frac{1}{\sqrt{2E(E+m)}}\begin{bmatrix}
			0\\t_1+t_2e^{ik_x}\\0\\-m-E
		\end{bmatrix}\label{Eq:ES_HE_Bot}\\
		&U^{\text{edge}}_{4,k_x}(y)=\left[1-\left(t_1/t_2\right)^2\right]^{1/2}e^{i\pi y}e^{(y-W)\ln\frac{t_2}{t_1}}\frac{1}{\sqrt{2E(E+m)}}\begin{bmatrix}
			-m-E\\0\\t_1+t_2e^{ik_x}\\0
		\end{bmatrix}
		\label{Eq:ES_HE_Top}
		\end{align}
	\end{subequations}

	Notice that the edge states localized at the bottom edge, Eq.~(\ref{Eq:ES_LE_Bot})(\ref{Eq:ES_HE_Bot}), are sublattice polarized to sublattice B and D, while those at the top edge, Eq.~(\ref{Eq:ES_LE_Top})(\ref{Eq:ES_HE_Top}) are polarized to sublattice A and C.
	
	

	\subsection{Quasi-1D geometry along $y$-direction}
	
	This subsection completely parallels the previous subsection.

	For the quasi-1D geometry infinite in $y$-direction and finite in $x$-direction, the quasi-momentum $k_y$ remains a good quantum number. $\kappa_x$ denotes transverse quantum number. The quasi-1D Bloch wavefunction now takes the form $\psi_{n/m,\kappa_x,k_y}(x,y)=e^{ik_yy}U_{n/m,\kappa_x,k_y}(x)$. The periodic part of the quasi-1D Bloch wavefunction takes the form of:
	\begin{equation}
		U_{n/m,\kappa_x,k_y}(x)=\alpha_1e^{i\kappa_xx}u_{1/3}(\kappa_x,k_y)+\alpha_2e^{-i\kappa_xx}u_{1/3}(-\kappa_x,k_y)		+\alpha_3e^{i\kappa_xx}u_{2/4}(\kappa_x,k_y)+\alpha_4e^{-i\kappa_xx}u_{2/4}(-\kappa_x,k_y)
		\label{Eq:Q1Dy_WF}
	\end{equation}
	where $\alpha_i$ with $i=1,2,3,4$ is the coefficient of superposition, to be determined from the boundary condition.
	
	
	The boundary condition is given by:
	\begin{subequations}
		\begin{align}
			&E_{n/m,\kappa_x}(k_y)U_{n/m,\kappa_x,k_y}(0)=h(k_y)U_{n/m,\kappa_x,k_y}(0)+T_x^{\dagger}U_{n/m,\kappa_x,k_y}(1)\label{Eq:yBC_Left} \\
		&E_{n/m,\kappa_x}(k_y)U_{n,\kappa_x,k_y}(W)=h(k_y)U_{n/m,\kappa_x,k_y}(W)+T_xU_{n/m,\kappa_x,k_y}(W-1)
		\label{Eq:yBC_Right}
		\end{align}
	\label{Eq:yBC}
	\end{subequations}
	where $h(k_y)=H_{\text{os}}+T_ye^{-ik_y}+T_y^{\dagger}e^{ik_y}$. The quasi-1D spectrum is given by:
	\begin{equation}
		E_{n/m,\kappa_x}(k_y)=\pm\sqrt{m^2+2t_1^2+2t_2^2+2t_1t_2\cos \kappa_x+2t_1t_2\cos k_y}.
	\end{equation}
	The boundary conditions in Eq.~(\ref{Eq:yBC_Left}-\ref{Eq:yBC_Right}) determines the combination coefficients $\alpha_i$ and the possible values of $\kappa_x$ in Eq.~(\ref{Eq:Q1Dy_WF}).
	
	For the extended states, if we want to study the left edge, then the boundary condition Eq.~(\ref{Eq:yBC_Left}) suggests that the quasi-1D wavefunctions should take the following form: (i) For the lower filled subbands, the periodic part of the quasi-1D Bloch wavefunctions are:
	\begin{eqnarray}
		&U_{1,\kappa_x,k_y}(x)=\frac{e^{i\kappa_xx}}{\sqrt{2(W+1)}}\frac{t_2+t_1e^{i\kappa_x}}{t_2+t_1e^{-i\kappa_x}}u_1(\kappa_x,k_y)-\frac{e^{-i\kappa_xx}}{\sqrt{2(W+1)}}u_1(-\kappa_x,k_y)\\
		&U_{2,\kappa_x,k_y}(x)=\frac{e^{i\kappa_xx}}{\sqrt{2(W+1)}}e^{2i\kappa_x}u_2(\kappa_x,k_y)-\frac{e^{-i\kappa_xx}}{\sqrt{2(W+1)}}u_2(-\kappa_x,k_y)
	\end{eqnarray}
	(ii) For the upper empty subbands, the periodic part of the quasi-1D Bloch wavefunctions are:
	\begin{eqnarray}
		&U_{3,\kappa_x,k_y}(x)=\frac{e^{i\kappa_xx}}{\sqrt{2(W+1)}}\frac{t_2+t_1e^{i\kappa_x}}{t_2+t_1e^{-i\kappa_x}}u_3(\kappa_x,k_y)-\frac{e^{-i\kappa_xx}}{\sqrt{2(W+1)}}u_3(-\kappa_x,k_y)\\
		&U_{4,\kappa_x,k_y}(x)=\frac{e^{i\kappa_xx}}{\sqrt{2(W+1)}}e^{2i\kappa_x}u_4(\kappa_x,k_y)-\frac{e^{-i\kappa_xx}}{\sqrt{2(W+1)}}u_4(-\kappa_x,k_y)
	\end{eqnarray}
	The boudanry condition at the right edge, Eq.~(\ref{Eq:yBC_Right}), furthur gives the quantization condition for the transverse quantum number $\kappa_x$:
	\begin{equation}
		\frac{t_2+t_1e^{-i\kappa_x}}{t_2+t_1e^{i\kappa_x}}=e^{2i\kappa_x(W+1)}.
		\label{Eq:QC1}
	\end{equation}
	For a wide sample $W\gg1$, the transverse quantum number $\kappa_x$ are very densely valued, with an average distance of $\pi/(W+1)$. Thus, practically, the summation over the transverse quantum number can be replaced by an integration.
	
	It turns out that the quantization condition, Eq.~(\ref{Eq:QC1}), has complex solution when $t_1<t_2$. In this case, the transverse quantum number may take the following value:
	\begin{equation}
		\kappa^c_x=\pi+i\ln\frac{t_2}{t_1}\ \ \ \ \ \text{for}\ (W+1)\ln\frac{t_2}{t_1}\gg1
	\end{equation}
	This solution corresponds to the gapped edge states.
	
	There are two degenerate edge bands with energy:
	\begin{equation}
		E^-_{\text{edge}}(k_y)=-E=-\sqrt{m^2+t_1^2+t_2^2+2t_1t_2\cos k_y}
	\end{equation}
	The corresponding wavefunctions are the following:
	\begin{subequations}
		\begin{align}
			&U^{\text{edge}}_{1,k_y}(x)=\left[1-\left(t_1/t_2\right)^2\right]^{1/2}e^{i\pi x}e^{-x\ln\frac{t_2}{t_1}}\frac{1}{\sqrt{2E(E+m)}}\begin{bmatrix}
			-t_1-t_2e^{ik_y}\\m+E\\0\\0
		\end{bmatrix} \label{Eq:ES_LE_Left}\\
		&U^{\text{edge}}_{2,k_y}(x)=\left[1-\left(t_1/t_2\right)^2\right]^{1/2}e^{i\pi x}e^{(x-W)\ln\frac{t_2}{t_1}}\frac{1}{\sqrt{2E(E+m)}}\begin{bmatrix}
			0\\0\\m+E\\t_1+t_2e^{-ik_y}
		\end{bmatrix}
		\label{Eq:ES_LE_Right}
		\end{align}
	\end{subequations}
	
	In addition, there are two degenerate edge bands with energy:
	\begin{equation}
		E^+_{\text{edge}}(k_y)=E=\sqrt{m^2+t_1^2+t_2^2+2t_1t_2\cos k_y}
	\end{equation}
	The corresponding wavefunctions are:
	\begin{subequations}
		\begin{align}
			&U^{\text{edge}}_{3,k_y}(x)=\left[1-\left(t_1/t_2\right)^2\right]^{1/2}e^{i\pi x}e^{-x\ln\frac{t_2}{t_1}}\frac{1}{\sqrt{2E(E+m)}}\begin{bmatrix}
			-m-E\\t_1+t_2e^{-ik_y}\\0\\0
		\end{bmatrix}  \label{Eq:ES_HE_Left}\\
		&U^{\text{edge}}_{4,k_y}(x)=\left[1-\left(t_1/t_2\right)^2\right]^{1/2}e^{i\pi x}e^{(x-W)\ln\frac{t_2}{t_1}}\frac{1}{\sqrt{2E(E+m)}}\begin{bmatrix}
			0\\0\\t_1+t_2e^{ik_y}\\-m-E
		\end{bmatrix}
		\label{Eq:ES_HE_Right}
		\end{align}
	\end{subequations}
	
	Notice that the edge states localized at the left edge, Eq.~(\ref{Eq:ES_LE_Left})(\ref{Eq:ES_HE_Left}), are sublattice polarized to sublattice A and B, while those at the right edge, Eq.~(\ref{Eq:ES_LE_Right})(\ref{Eq:ES_HE_Right}) are polarized to sublattice C and D.
	
	
	
	\twocolumngrid

	
	
	

	\bibliography{CCCD_ref}{}	
	
\end{document}